\documentclass{jaa}
\usepackage{graphicx}
\usepackage{xcolor, soul}

\begin{document}\sloppy

\title{Luminosity dependent cyclotron line in Swift J1626.6-5156}


\author{Binay Rai\textsuperscript{1}, Biswajit Paul\textsuperscript{2}, Mohammed Tobrej\textsuperscript{1}, Manoj Ghising\textsuperscript{1}, Ruchi Tamang\textsuperscript{1} and Bikash Chandra Paul\textsuperscript{1}}
\affilOne{\textsuperscript{1}Department of Physics, North Bengal University,Siliguri, Darjeeling, WB, 734013, India\\}
\affilTwo{\textsuperscript{2}ICARD, Depatment of Physics, North Bengal University, Siliguri, Darjeeling, WB, 734013, India}


\twocolumn[{

\maketitle

\corres{bcpaul@associates.iucaa.in}


\begin{abstract}
We studied the timing and spectral properties of the Be/X-ray pulsar Swift J1626.6-5156 using the \emph{NICER} observations of its 2021 outburst. The most important observation is the positive correlation of the centroid energy of the fundamental cyclotron line with the luminosity. This observation agrees with the usual positive correlation of the centroid energy cyclotron line with luminosity in the sub-critical regime. The correlation between the two quantities is verified using two different continuum models. The photon index decreases with an increase in flux. Thus, the spectrum is softer when the flux is low, which may be due to a decrease in the optical depth of the accretion column with a decrease in the flux. 
\end{abstract}


}]


\doinum{12.3456/s78910-011-012-3}
\artcitid{\#\#\#\#}
\volnum{000}
\year{0000}
\pgrange{1--}
\setcounter{page}{1}
\lp{1}

\section{Introduction}
Swift J1626.6-5156 is a $\sim$15 s Be/X-ray pulsar (Palmer 2005). It was discovered in 2005 during a giant Type II outburst by \textit{Swift} Burst Alert Telescope (BAT) (Krimm 2005) and went to a quiescent state in 2009 (Decesar 2013). The companion of the pulsar is B0Ve star. Reig (2011) estimated the distance to the pulsar's companion to be 10 kpc, but this distance is associated with significant uncertainty. The source distance estimated by Gaia EDR3 lies in the 5.8-12 kpc (Bailer-Jones et al. 2021). The 2005 outburst of the source decayed exponentially with flaring episodes (Reig 2008). A weak QPO at 1 Hz with fractional rms of 4.7 per cent was also reported by Reig et al. (2008). A long-term variability of $\sim$ 45 to 95 days is observed in the light curve of the source (DeCesar 2009; Baykal 2010). The binary orbit of the system is nearly circular with an orbital period of 132.9 days (Baykal 2010). Icdem et al. (2011) observed that the pulsar's spin-up rate and X-ray flux are correlated. This correlation was used to estimate the distance and magnetic field of the pulsar to be $\sim$15 kpc and $\sim$9 × 10$^{11}$ G, respectively (Icdem et al. (2011)). The spectral analysis performed by Icdem et al. (2011) revealed that the photon index exhibits an increasing trend with decreasing flux while the hydrogen column density decreases with a decrease in flux.

Cyclotron resonance scattering feature (CSRF) or cyclotron line has been observed in X-ray pulsars over a wide range of energies. The line energies have been estimated in the range $\sim$ 5-100 keV. In some sources, multiple cyclotron lines have been reported in the X-ray spectra (Heindl et al. 2004; Jaisawal et al. 2015; Molkov et al. 2021b; Pottschmidt et al. 2005). These lines are thought to be produced close to the neutron star, near the magnetic poles where the magnetic field is very strong. Electrons in the accreted matter are quantized in discrete energy levels (Landau levels) in a strong magnetic field. The cyclotron line is formed due to resonant scattering of the photons by these quantized electrons. This feature or line is essential because it provides a direct estimate of the pulsar's magnetic field. The centroid energy $E_{cyc}$ of the cyclotron line is written as (Staubert et al. 2019) $E_{cyc}\approx\dfrac{n}{1+z}11.6B_{12}(keV)$, where $B_{12}$ is the magnetic field in the order of 10$^{12}$ G, $z$ is the gravitational redshift due to the neutron star, $n$ denotes the discrete Landau levels.

DeCesar et al. (2013) confirmed the absorption-like feature at $\sim$ 10 keV in Swift J1626.6-5156 spectrum to be a cyclotron line. They observed that the energy of the absorption line shows a positive correlation with the source luminosity. The possible presence of the cyclotron line at 10 keV and its harmonic at 20 keV was also reported earlier by Coburn (2006). The recent 2021 outburst from the pulsar was first detected by MAXI/GSC (Negoro et al. 2021). The presence of prominent absorption features near 9 and 17 keV in \textit{NuSTAR} spectrum was observed by (Molkov et al. 2021a). Molkov et al. (2021b) reported the discovery of the cyclotron line at 5 keV, along with three others. Therefore, this source may be identified as the second source after 4U 0115+63 where more than three cyclotron lines have been detected.

 The centroid energy of the cyclotron line is found to depend on the luminosity of the X-ray pulsars. The first such dependence was observed in the 4U 0115+63, V 0332+53 and Cep X-4 (Makishima et al. 1990; Mihara et al. 1995), where $E_{cyc}$ showed anticorrelation with luminosity $(L_{x})$. The anticorrelation between $E_{cyc}-L_{x}$ in V 0332+53 was further confirmed by Tsygankov et al. (2006). The anticorrelation in 4U 0115+63 was due to an artifact of the continuum model used (Muller et al. 2013; Iyer et al. 2015). In case Cep X-4, the $E_{cyc}$ and $L_{x}$ were found to show a positive correlation instead of negative (Vybornov et al. 2017). The anticorrelation of $E_{cyc}-L_{x}$  was also observed in SMC X-2 (Jaisawal et al. 2016) and recently in 1A 0535+262 (Mandal et al. 2022). The sources which show positive $E_{cyc}-L_{x}$ correlations are Her X-1 (Staubert et al. 2007), GX 304-1 (Rothschild et al. 2017), Swift J1626.6-5156 (Decesar et al. 2013) and 2S 1553-542 (Malacaria et al. 2022). Doroshenko et al (2017) reported a positive correlation as well as anticorrelation of $E_{cyc}$ with $L_{x}$ in the case of V0332+53. The positive and negative correlations were observed in low and high luminosity, respectively. In other X-ray pulsars, the positive correlation is seen in the low luminosity state when the mass accretion rate is low, and negative correlation is seen in the high luminosity state where the accretion rate is high. Therefore, the same source may exhibit both positive and negative $E_{cyc}$-$L_{x}$ correlation depending upon their accretion state (Cusumano et al. 2016). Recently, the positive and negative correlations between $E_{cyc}$-$L_{x}$ in GRO J1008-57 were reported by Chen et al. (2021).

\section{Observation and Data Reduction}
The \texttt{Neutron Star Interior Composition Explorer} (\emph{NICER}) is an X-ray telescope onboard International Space Station (ISS) (Arzoumanian et al. 2014; Gendreau et al. 2014). It has 56 aligned Focal Plane Modules (FPMs) out of which 52 are currently operational, the two detectors (14,34) show an increase in detector noise. Each FPM consists of a detector package - detector, preamplifier, and thermoelectric cooler. It operates in the energy 0.2-12 keV energy range and has a time resolution of 100 ns. The energy resolution of the \emph{NICER} is about 150 eV at 6.5 keV.

 The data reduction and analysis were done using \textsc{HEASOFT} v6.30.1. The standard screening, filtering, and calibration of \emph{NICER} were performed using the tool \textsc{nicerl2} of the software \textsc{NICERDAS} version 9 using the latest calibration file xti20210720. The source spectra and the light curves were generated in \textsc{xselect}. The background spectra were obtained using the background estimating tool \textsc{nibackgen3c50} (Remillard et al. 2022). The model consists of three parameters, of which two track the different types of background events, and the third one predicts the excess count rate in the low energy due to observations done in the presence of sunlight. An extra model parameter is required in the high raw background rate in the 0.4-12 keV energy range (Remillard et al. 2022). \emph{NICER} uses the RXTE blank sky "BKGD$\_$RXTE$\#$" where $\#$ is a number between 1 to 8 excluding 7. In the prelaunch analysis, it was found that the \emph{NICER} background would be due to cosmic diffused X-ray background (which is very small due to its very small field of view (31.6 sq arcmin)) and particle interactions. The 3C50 model covers these two sources of the background. The prelaunch analysis showed that the background is 0.2 ct s$^{-1}$ in the 0.4-2.0 keV energy range and 0.15 ct s$^{-1}$ in 2-8 keV. The median background count rate estimated in the energy range 0.4-12 keV, obtained from 3556 good time intervals having an average duration of 570 s is 0.87 counts s$^{-1}$ making use of \textsc{nibackgen3c50} (Remillard et al. 2022).

\begin{table}
\begin{tabular}{c c c}
\hline				
Obs ID	&	Time	&	Exposure	\\
	&	(YYYY-MM-DD hh:mm:ss)	&	(ks)	\\
\hline	
& \emph{NICER} & \\
\hline					
4202070101	&	2021-03-11 16:35:29	&	2.8	\\
4202070102	&	2021-03-14 00:52:40	&	2.2	\\
4202070103	&	2021-03-14 23:39:30	&	3.4	\\
4202070104	&	2021-03-17 02:53:01	&	2.2	\\
4202070106	&	2021-03-19 09:04:40	&	3.9	\\
4202070107	&	2021-03-25 11:56:40	&	4.3	\\
4202070108	&	2021-03-26 03:34:40	&	0.8	\\
4202070109	&	2021-03-29 16:34:20	&	1.2	\\
4202070110	&	2021-03-30 17:21:24	&	1.2	\\
4202070111	&	2021-03-31 19:41:40	&	1.2	\\
4202070112	&	2021-04-01 18:57:17	&	1.2	\\
4202070114	&	2021-04-03 17:23:20	&	2.4	\\
4202070116	&	2021-04-05 00:23:36	&	2.3	\\
4202070117	&	2021-04-06 04:18:21	&	3.5	\\
4202070119	&	2021-04-08 09:03:20	&	3.6	\\
4202070159	&	2021-06-26 03:43:59	&	1.4	\\
\hline
& \emph{NuSTAR} &\\
\hline
90701311002 &2021-03-13 06:51:09 & 50.4\\
\hline					
\end{tabular}
\caption{Observation IDs of the observations used in the study along with the date of observations and exposures.}
\label{table-1}
\end{table} 
 
  The background light curves of the \emph{NICER} observations were extracted using another \emph{NICER} background estimation tool \textsc{nicer$\_$bkg$\_$estimator}(v 0p6, K. C. Gendreau \textit{et al.} 2022, in preparation). This alternate background estimator tool is based on the space weather method and uses the local spacecraft environment to generate background. The model uses local cutoff rigidity \textsc{COR$\_$SAX}, which measures the shielding provided by Earth's magnetic field, \textsc{K$_{p}$} index measures the magnetic disturbances and \textsc{SUN$\_$ANGLE}. The background estimated using this model is the instrumental background and does not include cosmic X-ray background along the direction of the source but instead uses cosmic X-ray background from the \emph{NICER} blank fields. A combination of the above three parameters may exist, which are not contained in the background events file, following which these times are not included in the resulting background spectrum. Other parameters may also exist along with the above three for better estimations of the background spectrum. Background light curves were generated using the Python module of the given background tool. The median value of the background count rate in 0.7-10 keV energy range obtained using \textsc{nicer$\_$bkg$\_$estimator} is 0.8 counts s$^{-1}$ for the source. The background correction of the light curves was done using the task \textsc{lcmath}. The light curves were barycentrically corrected using the orbit file with the help of \textsc{barycorr}. The response matrix file \textsc{rmf} for \emph{NICER} was generated using the tool \textsc{nicerrmf} and the corresponding ancillary response file using \textsc{nicerarf}. The spectra were fitted in the energy range 0.7-10 keV, as a background above 10 keV dominated the spectra. In some observations, background domination in the spectra starts below 7 keV, whereas some observations had very short exposure due to only a few spectral bins. We have not considered these observations for spectral analysis. Out of 57 observations, we have found only 16 observations suitable for the spectral analysis in 0.7-10 keV energy range.
 
 In the case of the \emph{NuSTAR} observation, the standard screening and filtering of \textit{NuSTAR} data of the source was done using \textsc{nupipeline} v0.4.8. A circular region of radius 100$^{\prime\prime}$ centered around the source was considered as the source extraction region, and another circular region of the same radius in the source free region was taken as the background extraction region. The source and background spectra were generated using these regions with the help of \textsf{nuproducts}. The observation IDs of the \emph{NICER} and \emph{NuSTAR} observations used here in the analysis are shown in Table \ref{table-1}. Spectra were fitted in X-ray spectral fitting software \textsc{xspec} v12.12.1 (Arnaud 1996).

\begin{figure*}
\begin{minipage}{0.3\textwidth}
\includegraphics[height=1.2\columnwidth]{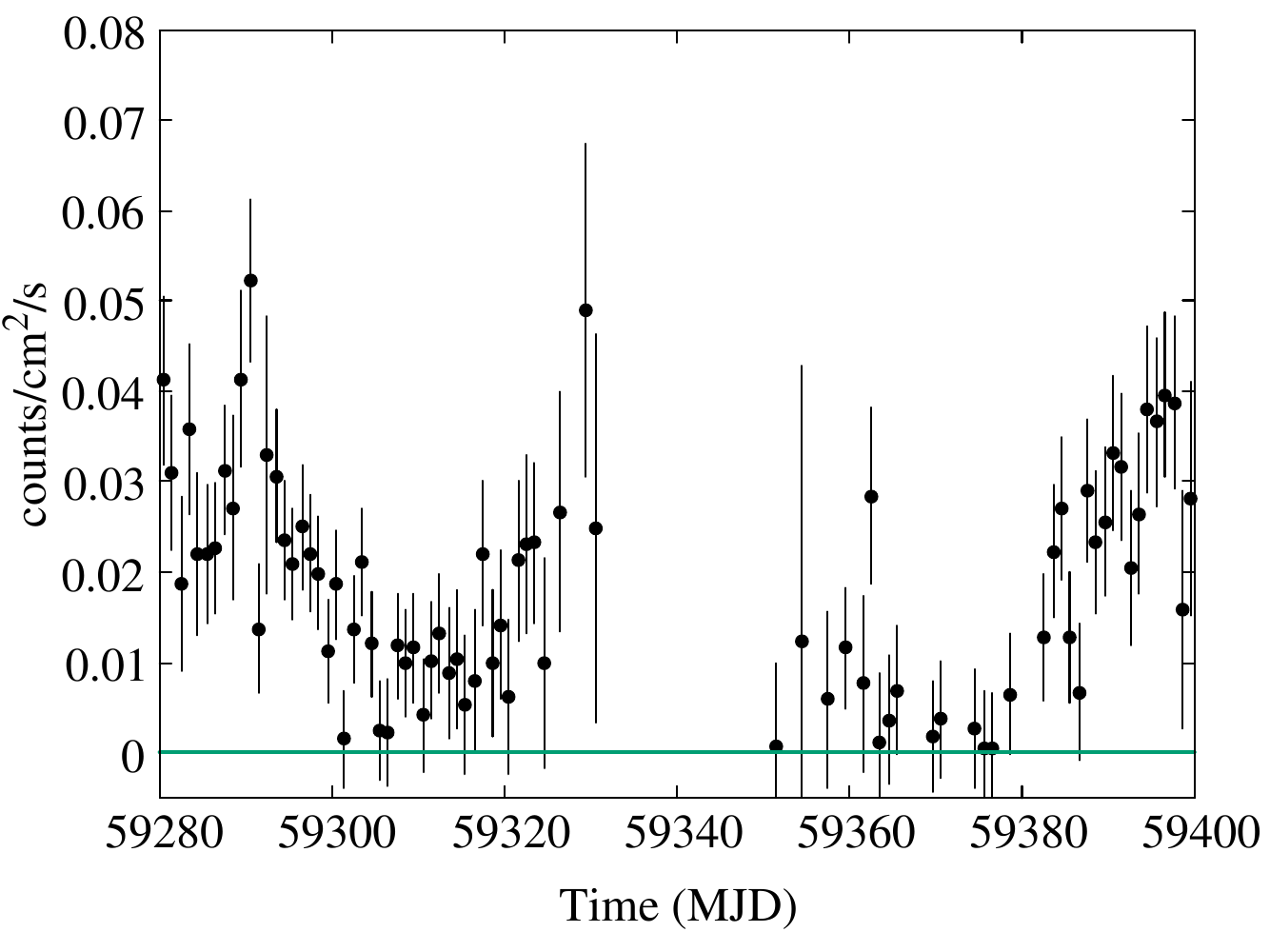}
\end{minipage}
\hspace{0.23\linewidth}
\begin{minipage}{0.3\textwidth}
\includegraphics[height=1.2\columnwidth]{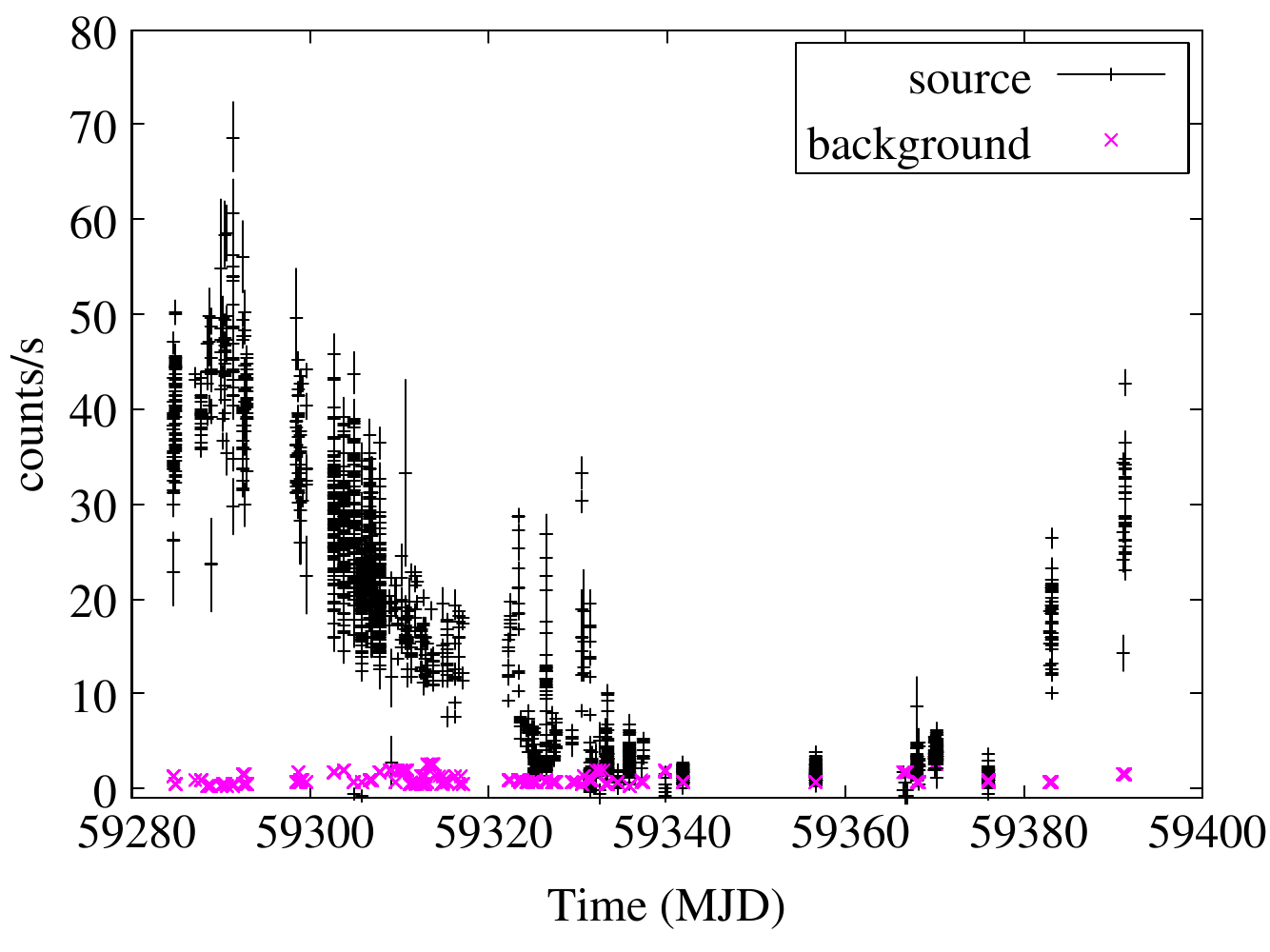}
\end{minipage}
\caption{\textit{Left} - Variation of the \emph{MAXI-GSC} (2-20 keV) count rate of Swift J1626.6-5156 with time. \textit{Right} -  \emph{NICER} (0.7-10 keV) light curve of source. The black points represents source count rates, whereas the magenta points indicates the background count rates.}
\label{Figure1}
\end{figure*}

\section{Timing Analysis}

\subsection{Light curves}

The \emph{MAXI} (2-20 keV) light curve of the Swift J1626.6-5156 along with that of the \emph{NICER} (0.7-10 keV) during 2021 outburst are shown on the left and right sides Figure \ref{Figure1} respectively. From the light curves we can observe that the source count rate is maximum at $\sim$ 59290 MJD and then decreases and continuously until 59320 MJD. After 59320 MJD the count rate increases followed by a decrease again. In between 59340-59389 MJD the count rates are extremly low and the pulsar may have entered the quiescent phase. After 59389 MJD there is increase in the source count rate.

\subsection{Pulse profiles}

\begin{figure}
\centering
\includegraphics[scale=0.35]{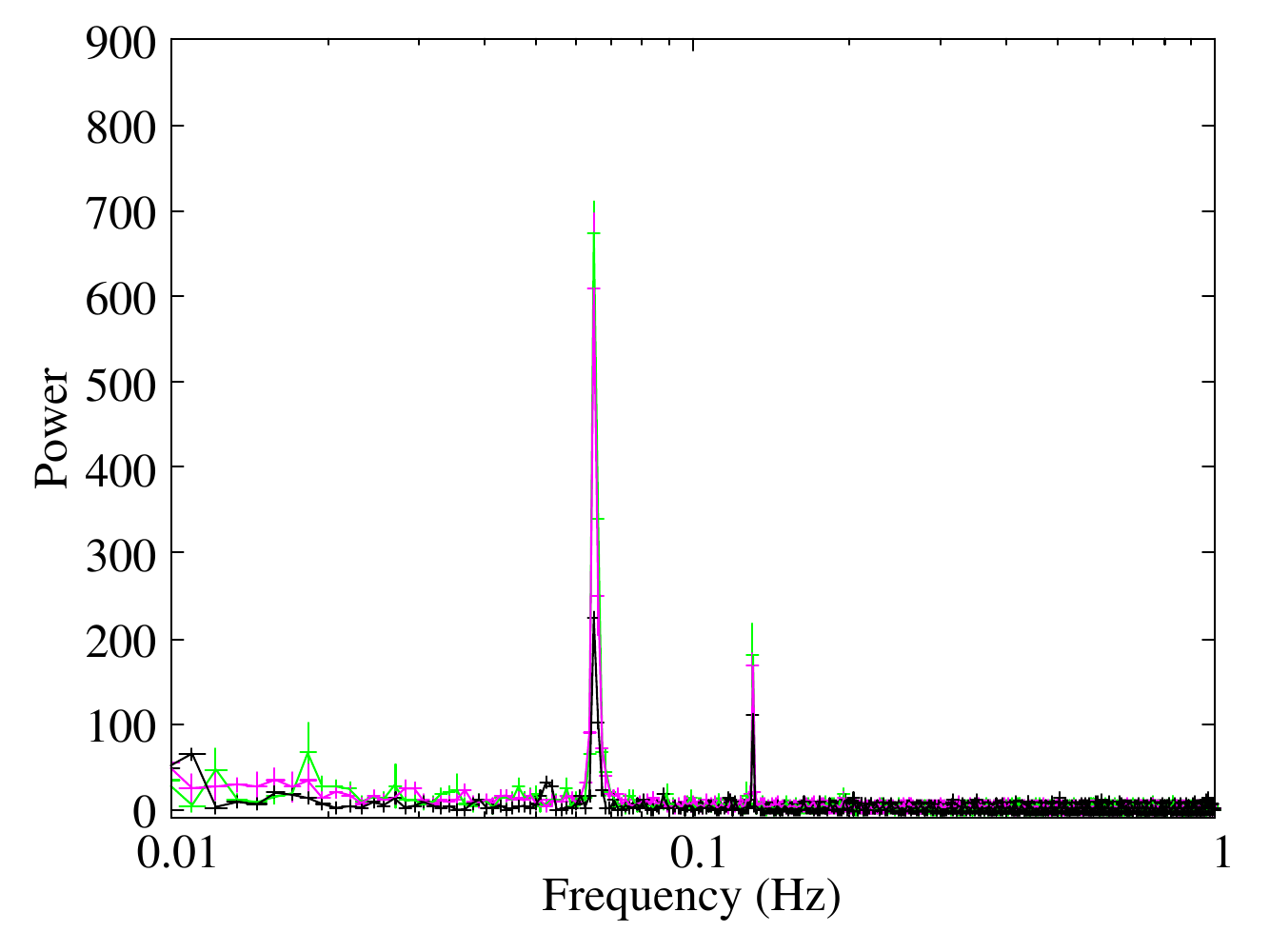}
\caption{The power spectra of the pulsar for three observations - 4202070101 (green), 4202070103 (magenta) and 4202070119 (black). The peaks in the power spectrum at $\sim$65 mHz and $\sim$ 130 mHz corresponds to fundamental and harmonic of the pulse period.}
\label{fig-2}
\end{figure}

\begin{figure}
\centering
\includegraphics[height=9 cm, width=12 cm, angle=-90]{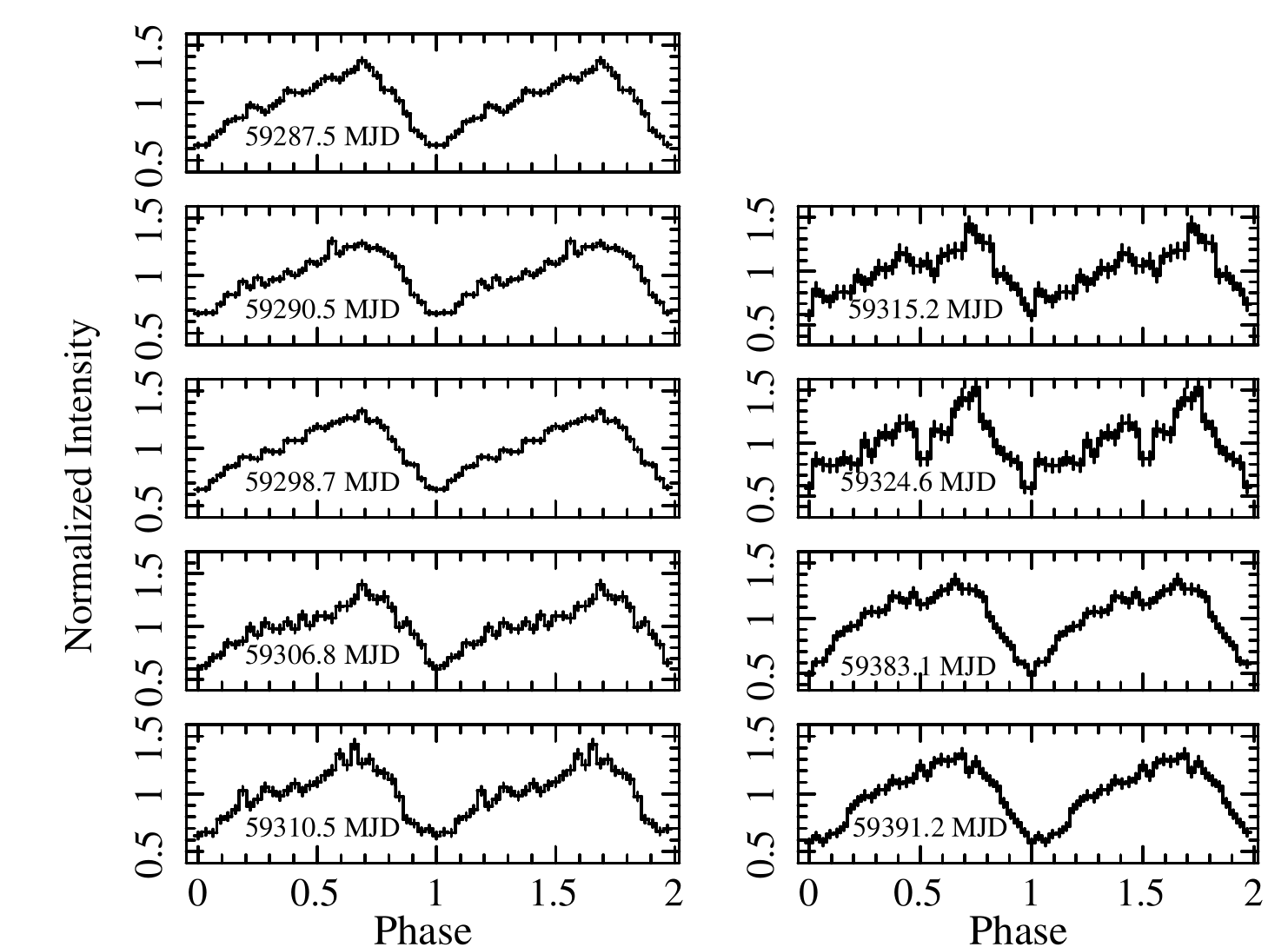}
\caption{Variation of the pulse profile of the pulsar with the time. The time quoted in the figure is the average time of the start and stop times of the observations. The y-axis indicates normalized intensity. The pulse profiles are scaled in such a way that it fits in a single frame.}
\label{fig-3}
\end{figure}

The pulse period of the pulsar is crudely estimated using \textsc{powspec} and is further refined using the epoch folding technique with the help of the tool \textsc{efsearch}. The power spectra of the pulsar for three different observations are shown in Figure \ref{fig-2}. The pulse profiles of the source at different phases of the outburst are obtained by folding the light curves of time resolution of 0.01 s in the energy range 0.7-10 keV about the pulse period of the pulsar using the \textit{ftool} \textsc{efold}. Each pulse profile consists of 32 bins. The variation of the pulse profiles with the time is given in Figure \ref{fig-3}. The pulse profiles are asymmetric with the slow, steady rise in between phase 0 to $\sim$0.7 followed by an abrupt decrease in intensity between $\sim$0.7 to 1.0. 

The pulse profiles are almost similar in shape when the source intensity is high. However at low-intensity state, a noticeable change in the shape of the profile is observed, especially the one in 59315.2 MJD, where a dip at the phase $\sim$0.62 along with an additional broad second peak at $\sim$0.4.

\section{Spectral analysis}
\begin{figure}
\includegraphics[scale=0.3, angle=-90]{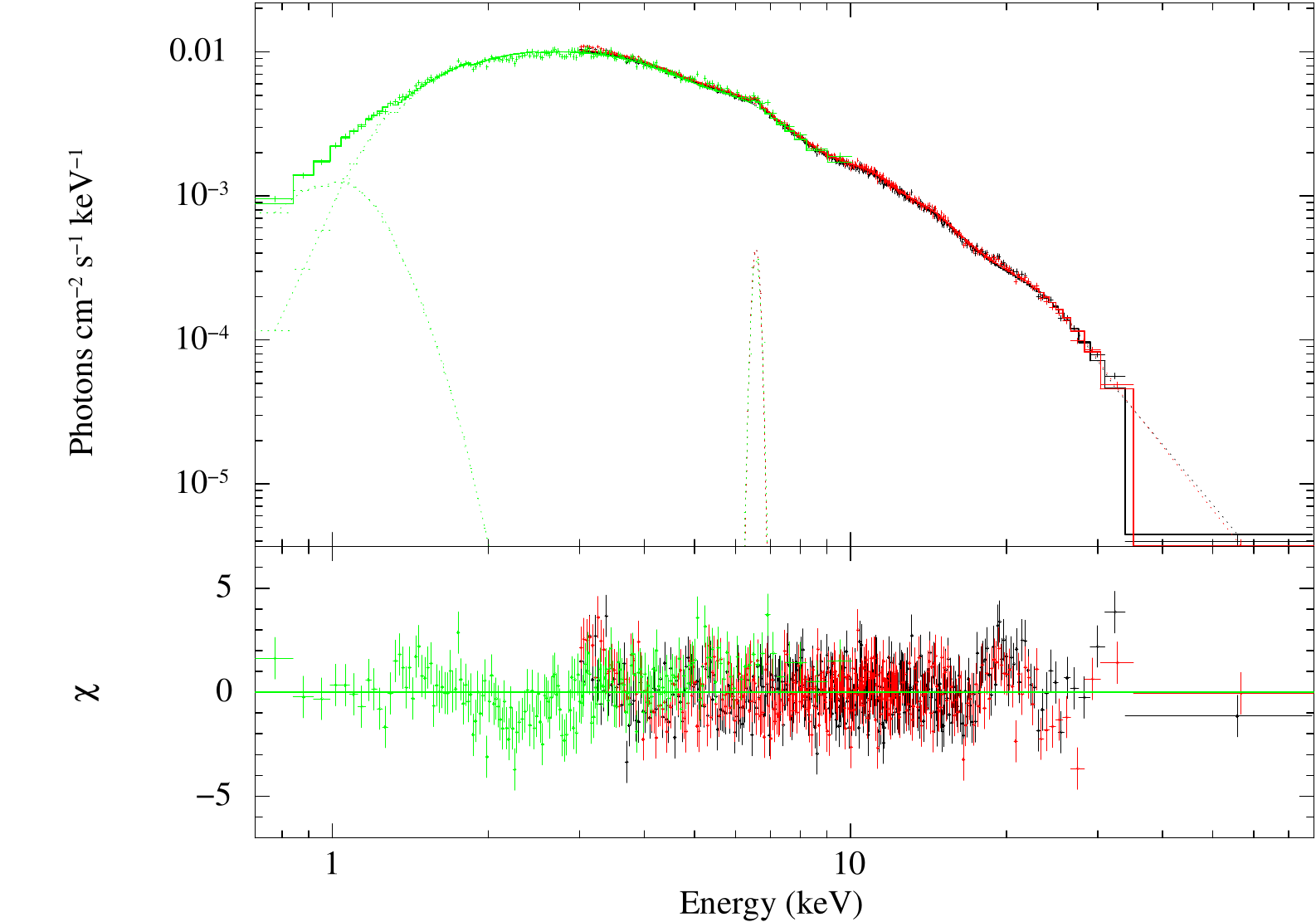}
\caption{The unfolded spectra of the simultaneously fitted \emph{NICER} and \emph{NuSTAR} spectra (upper panel). The bottom panel indicates the residuals of the fitting. The composite model used here is  \textsc{constant}$\times$\textsc{phabs}$\times$(\textsc{bbody}+\textsc{cutoffpl})$\times$\textsc{gabs}$\times$\textsc{gabs}$\times$\textsc{gabs}$\times$\textsc{gabs}.}
\label{fig-4}
\end{figure}

 The \emph{NICER} spectra of the source were grouped in such a way that each bin contains a minimum of 25 counts. We observed that the spectral parameters like plasma temperature ($kT$) and cutoff energy ($E_{cut}$) of the continuum models \textsc{cutoffpl} and \textsc{comptt} respectively obtained by fitting the \emph{NICER} spectra are small, which is unrealistic. These may be due to the narrow energy range of \emph{NICER} spectra. To make the \emph{NICER} spectral fitting more realistic, we first performed a broadband spectral fitting in 0.7-79 keV energy range using \emph{NuSTAR} and \emph{NICER} spectra and then used the values of $E_{cut}$ and $kT$ which fitting the \emph{NICER} spectra. The \emph{NuSTAR} and \emph{NICER} spectra were simultaneously fitted using \textsc{constant}$\times$\textsc{phabs}$\times$(\textsc{bbody}+\textsc{cutoffpl}+\textsc{gaussian})$\times$\textsc{gabs}$\times$\textsc{gabs}$\times$\textsc{gabs}$\times$\textsc{gabs} models. The model \textsc{gabs} is a Gaussian absorption model used to fit the cyclotron line in the spectra of X-ray pulsars. The four gabs models were used to fit the four cyclotron lines present at $\sim$ 5 keV, $\sim$ 9 keV, $\sim$ 12 keV, and $\sim$ 17 keV (Molkov et al. 2021b). However, we have fixed the centroid energy of the fourth harmonic cyclotron line to 17 keV as keeping this parameter free will set its values to 16 keV, which is inconsistent with the value reported by Molkov et al. 2021b. The broadband spectrum fits well by replacing the continuum model \textsc{cutoffpl} with another continuum model \textsc{comptt}. In the latter case, we found the centroid energy of the fourth harmonic cyclotron line to be 17.11 keV. The best-fit spectrum is shown in Figure \ref{fig-4}, and the best-fit values of the spectral parameters are tabulated in Table \ref{tab2}.

 After that, we fitted the \emph{NICER} spectra in 0.7-10 keV energy range with the composite model - \textsc{phabs}$\times$(\textsc{bbody}+\textsc{cutoffpl})$\times$\textsc{gabs}$\times$\textsc{gabs}. The last two \textsc{gabs} components used in the broadband spectral fitting are dropped while performing the \emph{NICER} spectral fitting, as the last two cyclotron lines were beyond the energy range over which spectral fitting is performed. The \textsc{gaussian} model was not required for fitting the \emph{NICER} spectra as there was no evidence of an iron line in the spectra. It is found that the spectral parameters of $\sim$ 9 keV cyclotron line in \emph{NICER} spectra cannot be constrained. The cut-off energy ($E_{cut}$) of the model \textsc{cutoffpl} was fixed to 5.71 keV, which is obtained from broadband spectral fitting as the cut-off energy obtained from the \emph{NICER} spectra are unrealistic due to its narrow energy range. The flux in the 0.7-10.0 keV energy range was extrapolated into bolometric flux in the energy range 0.1-100 keV. The photon index is found to lie between -0.01 to 0.52. The centroid energy of the $\sim$5 keV cyclotron line is found to vary between 4.72-5.26 keV with the luminosity varying between (1.82-10.18)$\times$10$^{36}$D$_{10}^{2}$ erg s$^{-1}$. Here D$_{10}$ denotes the distance in terms of 10 kpc. An absorption-like feature between 2.2-3.5 keV is also observed in the residuals of the fitted spectra. This feature can be due to the reflectivity of gold M shell.

\begin{table}
\centering
\scalebox{0.8}{
\begin{tabular}{c c c c}
\hline												
Model	&	Parameters		&		Values (\textsc{cutoffpl})		&		Values (\textsc{comptt})		\\
\hline												
\textsc{constant}	&	$C_{FPMA}$		&	1.00	(fixed)	&	1	(fixed)	\\
	&	$C_{FPMB}$		&	1.026	$\pm$	0.004	&	1.026	$\pm$	0.002	\\
	&	$C_{NICER}$		&	0.987	$\pm$	0.008	&	0.993	$\pm$	0.005	\\
\textsc{phabs}	&	$n_{H}$	(10$^{22}$ cm$^{-2}$)	&	1.11	$\pm$	0.04	&	0.45	$\pm$	0.04	\\
\textsc{bbody}	&	$kT$	(keV)	&	0.112	$\pm$	0.006	&	0.10	$\pm$	0.03	\\
\textsc{comptt}	&	$T_{0}$	(keV)	&	$\dots$			&	1.06	$\pm$	0.04	\\
	&	$kT$	(keV)	&	$\dots$			&	5.21	$\pm$	0.09	\\
	&	taup		&	$\dots$			&	4.03	$\pm$	0.14	\\
	&	norm		&	$\dots$			&	0.0167	$\pm$	0.0004	\\
\textsc{cutoffpl}	&	PhoIndex ($\alpha$)		&	0.18	$\pm$	0.06	&	$\dots$			\\
	&	$E_{cut}$	(keV)	&	5.71	$\pm$	0.16	&	$\dots$			\\
	&	norm		&	0.025	$\pm$	0.001	&	$\dots$			\\
\textsc{gaussian}	&	$E_{Fe}$	(keV)	&	6.58	$\pm$	0.03	&	6.58	$\pm$	0.03	\\
	&	$\sigma_{Fe}$	(keV)	&	0.12	$\pm$	0.04	&	0.13	$\pm$	0.04	\\
\textsc{gabs}	&	$E_{cyc_{1}}$	(keV)	&	5.05	$\pm$	0.04	&	4.90	$\pm$	0.03	\\
	&	$\sigma_{cyc_{1}}$	(keV)	&	0.75	$\pm$	0.05	&	0.79	$\pm$	0.09	\\
	&	$D_{cyc_{1}}$		&	0.26	$\pm$	0.04	&	0.26	$\pm$	0.08	\\
\textsc{gabs}	&	$E_{cyc_{2}}$	(keV)	&	8.89	$\pm$	0.05	&	8.77	$\pm$	0.06	\\
	&	$\sigma_{cyc_{2}}$	(keV)	&	1.86	$\pm$	0.07	&	1.44	$\pm$	0.13	\\
	&	$D_{cyc_{2}}$		&	2.45	$\pm$	0.17	&	0.90	$\pm$	0.22	\\
\textsc{gabs}	&	$E_{cyc_{3}}$	(keV)	&	12.65	$\pm$	1.30	&	12.75	$\pm$	0.11	\\
	&	$\sigma_{cyc_{3}}$	(keV)	&	1.16	$\pm$	1.80	&	0.99	$\pm$	0.13	\\
	&	$D_{cyc_{3}}$		&	0.53	$\pm$	22.54	&	0.26	$\pm$	0.06	\\
\textsc{gabs}	&	$E_{cyc_{4}}$	(keV)	&	17.00		(fixed)	&	17.11	$\pm$	0.10	\\
	&	$\sigma_{cyc_{4}}$	(keV)	&	3.89	$\pm$	0.18	&	1.63	$\pm$	0.16	\\
	&	$D_{cyc_{4}}$		&	5.07	$\pm$	0.49	&	0.78	$\pm$	0.12	\\
	& 	$\chi^{2}/dof$		&	2581.28	/	2141	&	2426.17	/2139		\\
\hline												
											
\end{tabular}
}
\caption{Best fit spectral parameters obtained through simultaneous fitting of the \textsc{NICER} and \textsc{NuSTAR} spectra in the broadband 0.7-79 keV energy range. The subscript 1, 2, 3 and 4 in the parameters of the \textsc{gabs} models represents the four absorption features observed at $\sim$ 5 keV, $\sim$ 9 keV, $\sim$ 12 keV and $\sim$ 17 keV respectively. The norm is expressed in the unit of photons/cm$^{2}$/s/keV at 1 keV. Errors quoted in the table are within 1$\sigma$ confidence interval.}
\label{tab2}
\end{table}

We have shown the values of best fit spectral parameters with and without \textsc{gabs} model required to fit $\sim$ 5 keV cyclotron line for three observations 4202070101, 4202070103 and 4202070119 in Table \ref{tab3}. From Table \ref{tab3}, it is evident that the addition of \textsc{gabs} model leads to a significant improvement in the fitting (decrease in the value of $\chi^{2}$). The observed change in $\chi^{2}$ ($\Delta\chi^{2}$) with and without \textsc{gabs} model in three observations are 68.9, 163.5  and 105.6 respectively. The check the statistical significance of the $\sim$ 5 keV absorption feature we simulated the 1000 spectra in \textsc{xspec} using the script \textsc{simftest}. From each simulated spectrum, the change in $\chi^{2}$ ($\Delta\chi^{2}$) with and without the addition of the \textsc{gabs} has been computed. The histogram of  $\Delta\chi^{2}$ follows $\chi^{2}$ distribution approximately with three degrees of freedom (Salganik et al. 2022). The maximum values of $\Delta\chi^{2}$ obtained from the simulation are 23.7, 20 and 23.5 for 4202070101, 4202070103 and 4202070119 respectively. These values are significantly lower than the observed values of $\Delta\chi^{2}$. The probabilities of the false detection of this feature are 7$\times$10$^{-15}$, 3$\times$10$^{-35}$ and 10$^{-22}$ respectively for three different observations. Thus, the significance of the feature in these cases is above $5\sigma$.

\begin{table*}
\begin{tabular}{ccccc}
\hline
& & \textsc{phabs}$\times$\textsc{(bbody+cutoffpl)}$\times$\textsc{gabs} & & \\
\hline
Models & Parameters & 4202070101 & 4202070103 &4202070119 \\

\hline
\textsc{phabs}	&	$n_{H}$	$(10^{22}	cm^{-2})$	&	1.14	$\pm$	0.05	&	1.1	$\pm$	0.04	&	1.18	$\pm$	0.06	\\
\textsc{bbody}	&	$kT$	(keV)		&	0.099	$\pm$	0.005	&	0.102	$\pm$	0.005	&	0.077	$\pm$	0.005	\\
	&	norm			&	0.005	$\pm$	0.002	&	0.003	$\pm$	0.001	&	0.014	$\pm$	0.008	\\
\textsc{cutoffpl}	&	$PhoIndex	(\alpha)$		&	0.02	$\pm$	0.03	&	0.08	$\pm$	0.02	&	0.72	$\pm$	0.04	\\
	&	$E_{cut}$	(keV)		&	5.71	(fixed)		&	5.71	(fixed)		&	5.71	(fixed)		\\
	&	norm			&	0.023	$\pm$	0.001	&	0.022	$\pm$	0.001	&	0.014	$\pm$	0.001	\\
\textsc{gabs}	&	$E_{cyc_{2}}$	(keV)		&	8.89	(fixed)		&	8.89	(fixed)		&	8.89	(fixed)		\\
	&	$\sigma_{cyc_{2}}$	(keV)		&	1.86	(fixed)		&	1.86	(fixed)		&	1.86	(fixed)		\\
	&	$D_{cyc_{2}}$			&	2.45	(fixed)		&	2.45	(fixed)		&	2.45	(fixed)		\\
																	
	&	$\chi^{2}/dof$	&	826.97/726	&	978.77/831	&	795.75/594	\\

\hline
& & \textsc{phabs}$\times$\textsc{(bbody+cutoffpl)}$\times$\textsc{gabs}$\times$\textsc{gabs} & & \\

\hline
 \textsc{phabs}	&	$n_{H}$	$(10^{22} cm^{-2})$	&	1.116	$\pm$	0.05	&	1.06	$\pm$	0.04	&	1.071	$\pm$	0.064	\\
\textsc{bbody}	&	$kT$	(keV)	&	0.101	$\pm$	0.005	&	0.105	$\pm$	0.006	&	0.079	$\pm$	0.006	\\
	&	norm		&	0.004	$\pm$	0.001	&	0.002	$\pm$	0.001	&	0.007	$\pm$	0.005	\\
\textsc{cutoffpl}	&	$PhoIndex	(\alpha)$	&	0.09	$\pm$	0.03	&	0.001	$\pm$	0.025	&	0.521	$\pm$	0.056	\\
	&	$E_{cut}$	(keV)	&	5.17	$\pm$	(fixed)	&	5.17	$\pm$	(fixed)	&	5.17	$\pm$	(fixed)	\\
	&	norm		&	0.021	$\pm$	0.001	&	0.0201	$\pm$	0.0007	&	0.011	$\pm$	0.001	\\
\textsc{gabs}	&	$E_{cyc_{1}}$	(keV)	&	4.97	$\pm$	0.15	&	5.04	$\pm$	0.11	&	4.89	$\pm$	0.14	\\
	&	$\sigma_{cyc_{1}}$	(keV)	&	0.57	$\pm$	0.14	&	0.68	$\pm$	0.103	&	0.715	$\pm$	0.12	\\
	&	$D_{cyc_{1}}$		&	0.16	$\pm$	0.05	&	0.227	$\pm$	0.051	&	0.458	$\pm$	0.125	\\
\textsc{gabs}	&	$E_{cyc_{2}}$	(keV)	&	8.89	(fixed)		&	8.89	(fixed)		&	8.89		(fixed)	\\
	&	$\sigma_{cyc_{2}}$	(keV)	&	1.86	(fixed)		&	1.86	(fixed)		&	1.86		(fixed)	\\
	&	$D_{cyc_{2}}$		&	2.45	(fixed)		&	2.45	(fixed)		&	2.45		(fixed)	\\
	&	$\chi^{2}/dof$		&	758.01	/	723	&	815.3	/	828	&	690.15		/591	\\

\hline 

\end{tabular}
\caption{The best fitted spectral parameters with and without \textsc{gabs} model used to fit the $\sim$ 5 keV absorption feature for three different \emph{NICER} observations. The subscript 1 and 2 in the parameters of \textsc{gabs} models indicates the models parameters for $\sim$ 5 keV and $\sim$ 9 keV absorption features respectively. $D$ denotes the strength of the absorption features. The uncertainties associated with the spectral parameters are within 90\% confidence interval.}
\label{tab3}
\end{table*}

The second model combinations \textsc{phabs}$\times$(\textsc{bbody}+\textsc{comptt})$\times$\textsc{gabs}$\times$\textsc{gabs} also fits the \emph{NICER} spectra equally well. The model \textsc{comptt} is an analytic model that describes the soft photons' comptonization by the hot plasma (Titarchuk 1994). We have fixed the plasma temperature $kT$ of the model \textsc{comptt} to 5.21 keV, obtained from broadband spectral fitting. The soft photon temperature ($T_{0}$) and the optical depth parameters of the model were set free to vary. The spectral parameters of the $\sim$ 9 keV absorption feature were fixed to the value given in Table \ref{tab2}, when \textsc{comptt} is used as the continuum model. We have considered a disk-type geometry while using the model \textsc{comptt}.

\begin{figure*}
\centering
\includegraphics[height=0.8\columnwidth]{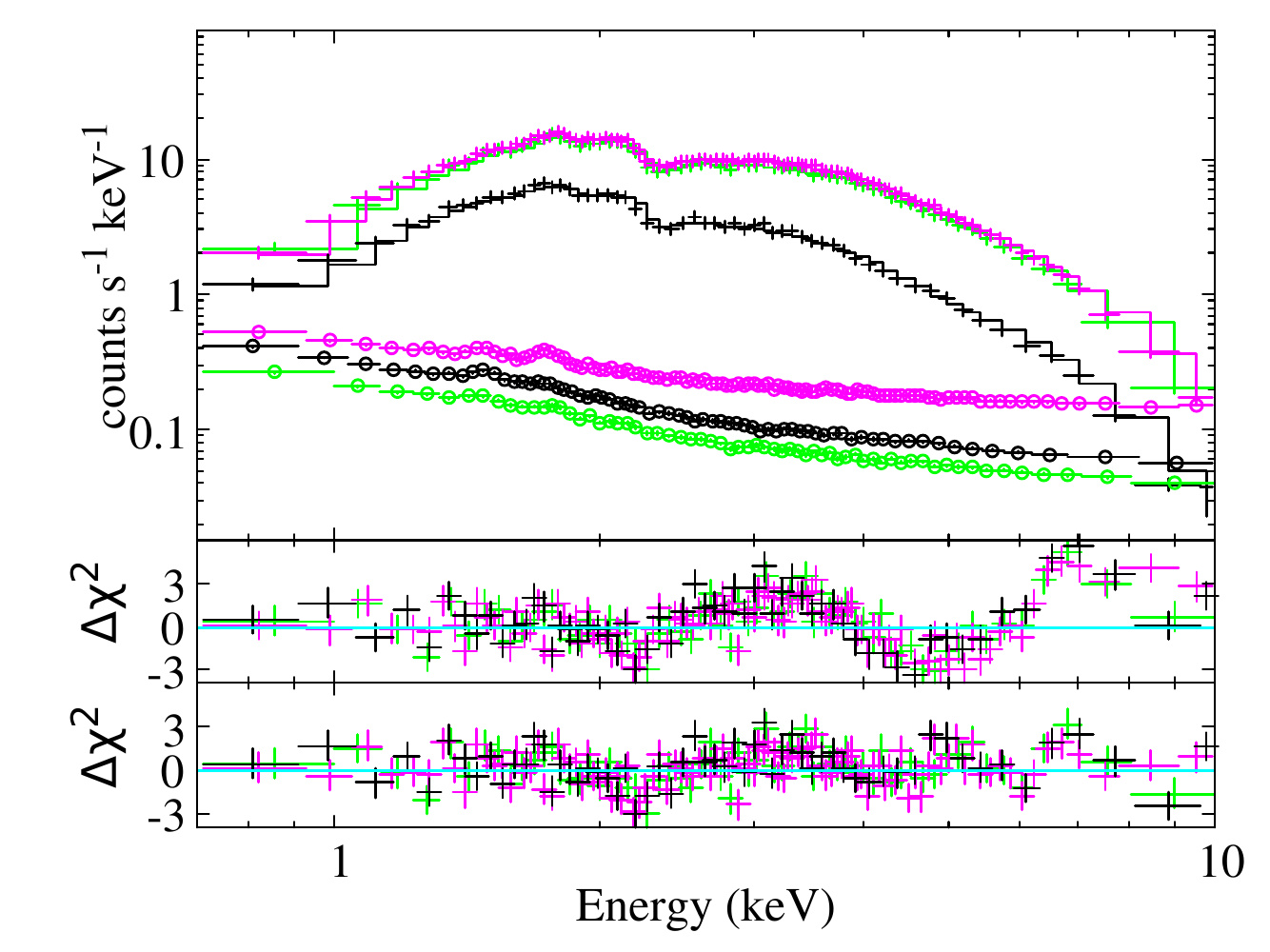}
\caption{Best fit spectra for three observations - 4202070101 (green), 4202070103 (magenta) and 4202070119 (black). The unfilled circles represents the corresponding background spectra.}
\label{fig-5}
\end{figure*}

\subsection{Variation of the $\sim$ 5 keV cyclotron line with luminosity}

\begin{figure*}
\begin{minipage}{0.3\textwidth}
\includegraphics[height=1.2\columnwidth]{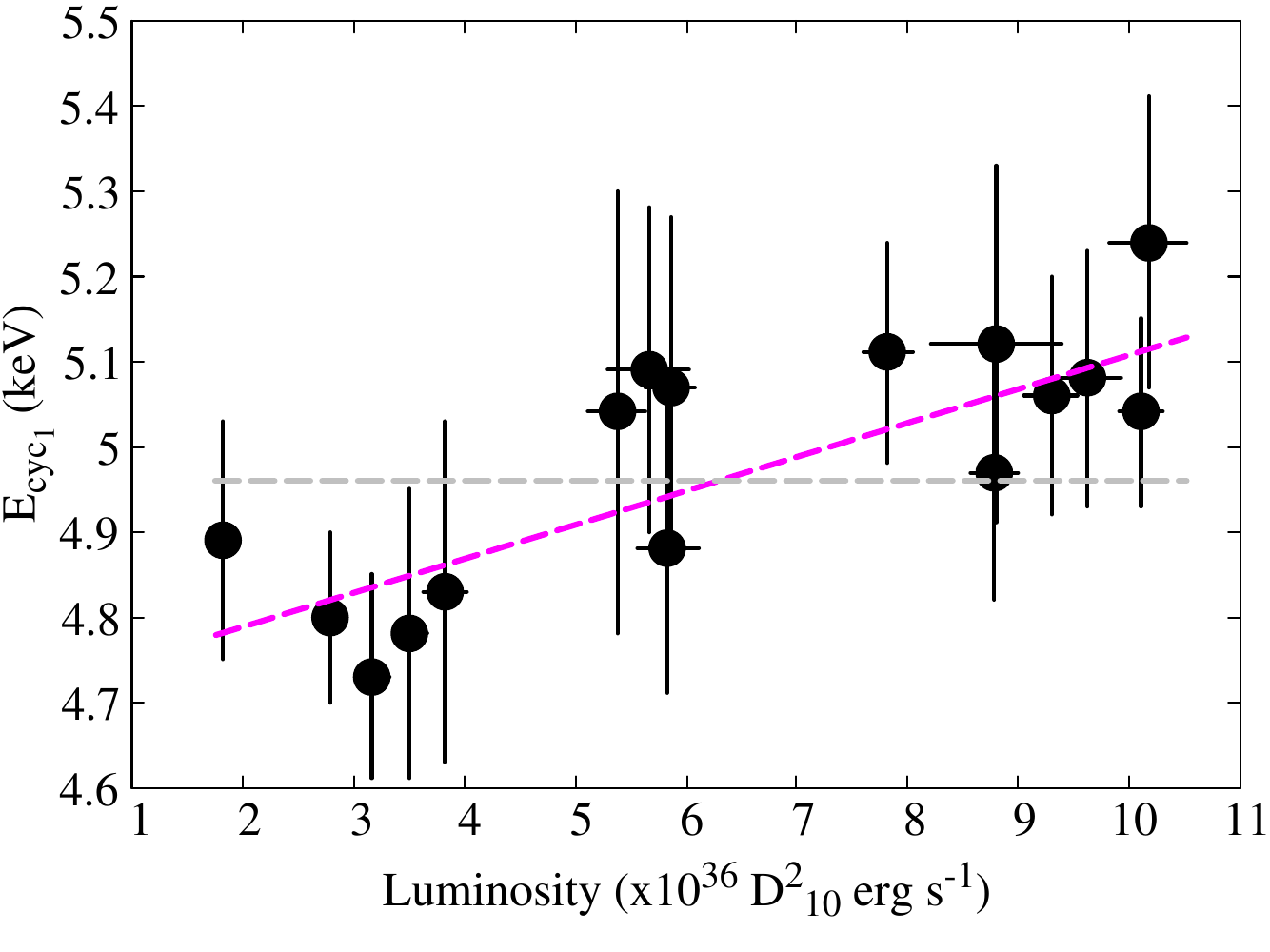}
\end{minipage}
\hspace{0.18\linewidth}
\begin{minipage}{0.3\textwidth}
\includegraphics[height=1.2\columnwidth]{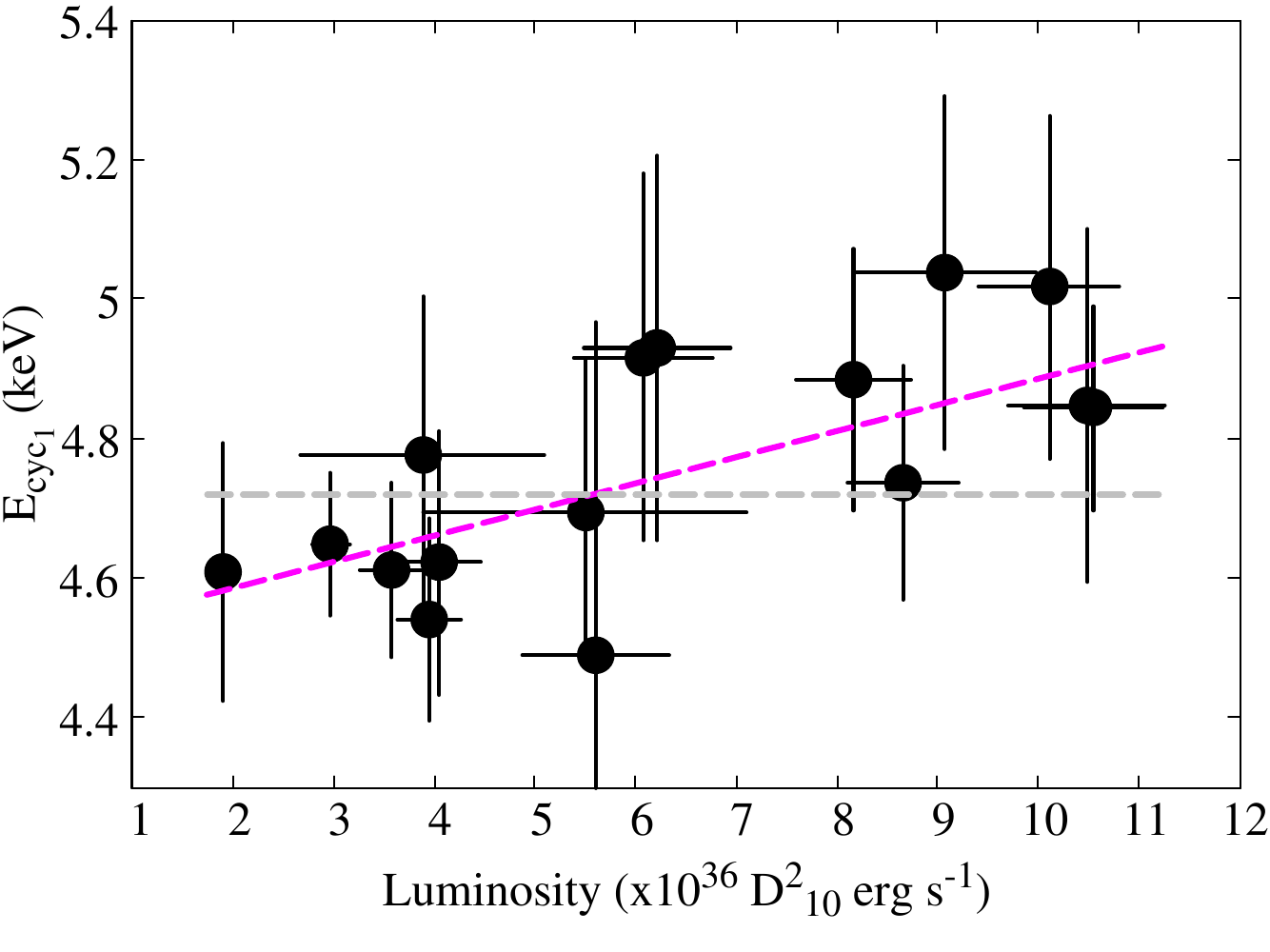}
\end{minipage}
\caption{\textit{Left - } Variation of $\sim$ 5 keV cyclotron line energy with luminosity in the case when \textsc{cutoffpl} is used as continuum model. \textit{Right -} Variation of the cyclotron line energy with luminosity for the continuum model \textsc{comptt}. The dashed lines in grey and magenta colours indicates the constant and linear models respectively.}
\label{fig-6}
\end{figure*}

The overall variation of the centroid energy of $\sim$ 5 keV cyclotron line ($E_{cyc_{1}}$) with luminosity ($L_{x}$), using two different continuum models has been presented in Figure \ref{fig-6}. Pearson's correlation coefficient is 0.78 when \textsc{cutoffpl} is used as continuum model, revealing a positive correlation between the line energy and luminosity. In order to explore the significance of the correlation, we fitted the variation of $E_{cyc_{1}}-L_{x}$ separately by a constant and linear model. The model that fits the variation of $E_{cyc_{1}}$ with $L_{x}$ well is determined with the help of the Bayesian Information Criterion (BIC). The BIC value for a constant model is 18.2 ($\chi^{2}=$15.42, 16 data points). However, in the case of the linear model, the BIC is 10.4 ($\chi^{2}=$4.89, 16 data points). The change in BIC ($\Delta BIC$) is $\sim$ 8, which gives positive evidence that a linear model fits the given variation between $E_{cyc_{1}}$ with $L_{x}$ well than a constant model. We have verified this using $F-test$. Considering a constant model to be our first model and the linear to be the second, we computed the $F$ value. The $F$-value obtained is about 30.159, and the corresponding p-value is $\sim$ 0.8$\times$10$^{-4}$. Therefore, a linear model is found to be statistically better (by more than $3\sigma$ level of significance) than a constant model, and this implies the existence of a correlation between $E_{cyc_{1}}$ and $L_{x}$.

To study whether this correlation of $E_{cyc_{1}}$ with $L_{x}$ is independent of the continuum model used, we plotted the variation of the $\sim$ 5 keV cyclotron line obtained using the continuum model \textsc{comptt}. We fitted it with a constant and linear model separately. The BIC for the constant model is $\sim$ 12 ($\chi^{2}=$9.62, 16 data points), and for the linear model, it is $\sim$ 9.3 ($\chi^{2}=$3.77, 16 data points). Thus $\Delta BIC$ is about 3, indicating positive evidence that the linear model is better than the constant model. The $F$ value computed using the constant and linear model is $\sim$ 21.7 with $p \sim$ 3.7$\times$10$^{-4}$. The correlation coefficient between the two quantities is 0.7. Thus, there exists a positive correlation between $E_{cyc_{1}}$ and $L_{x}$, and it is independent of the continuum models used in the fitting.

\begin{figure*}
\begin{minipage}{0.3\textwidth}
\includegraphics[height=1.2\columnwidth,angle=-90]{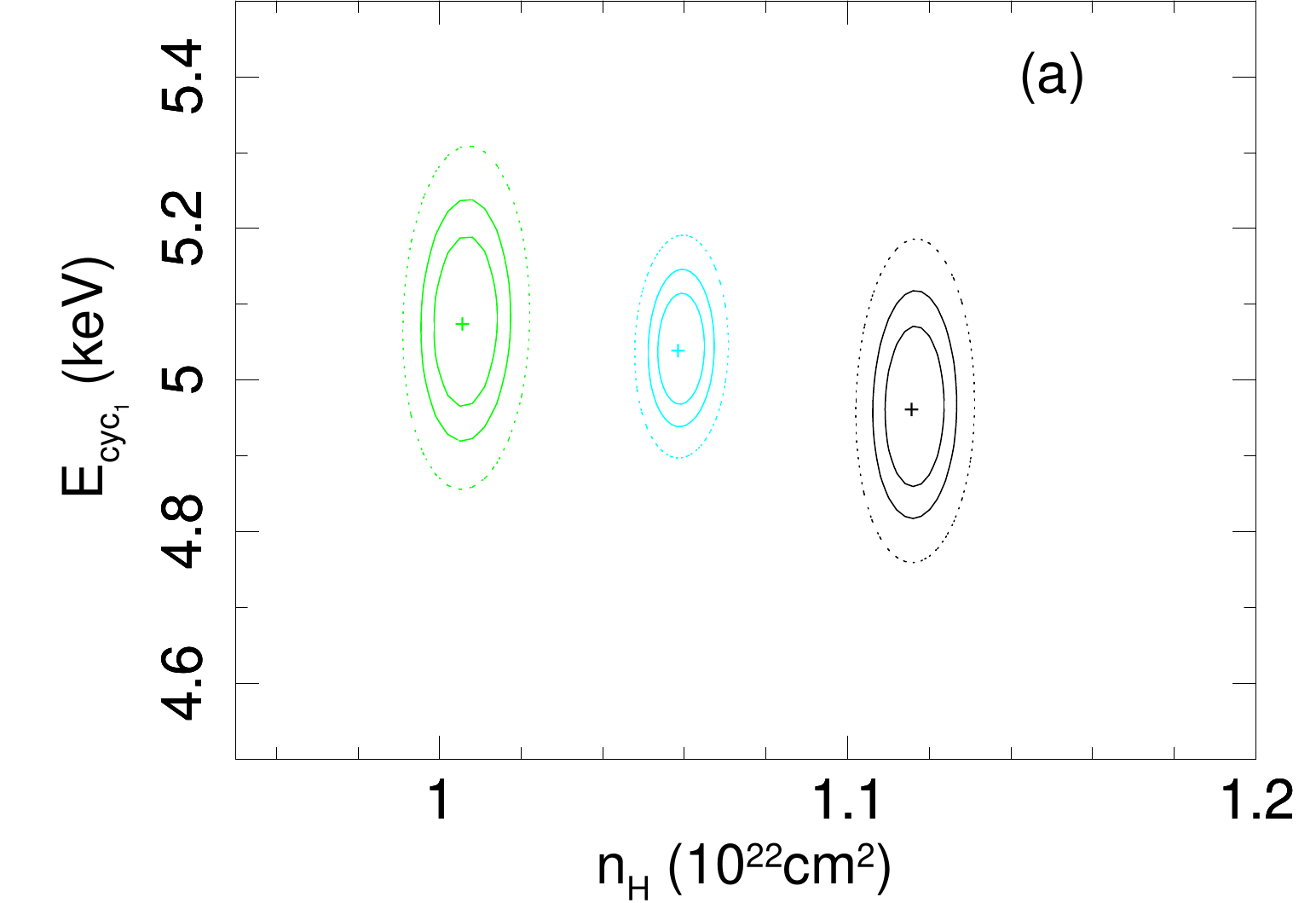}
\end{minipage}
\hspace{0.05\linewidth}
\begin{minipage}{0.3\textwidth}
\includegraphics[height=1.17\columnwidth,angle=-90]{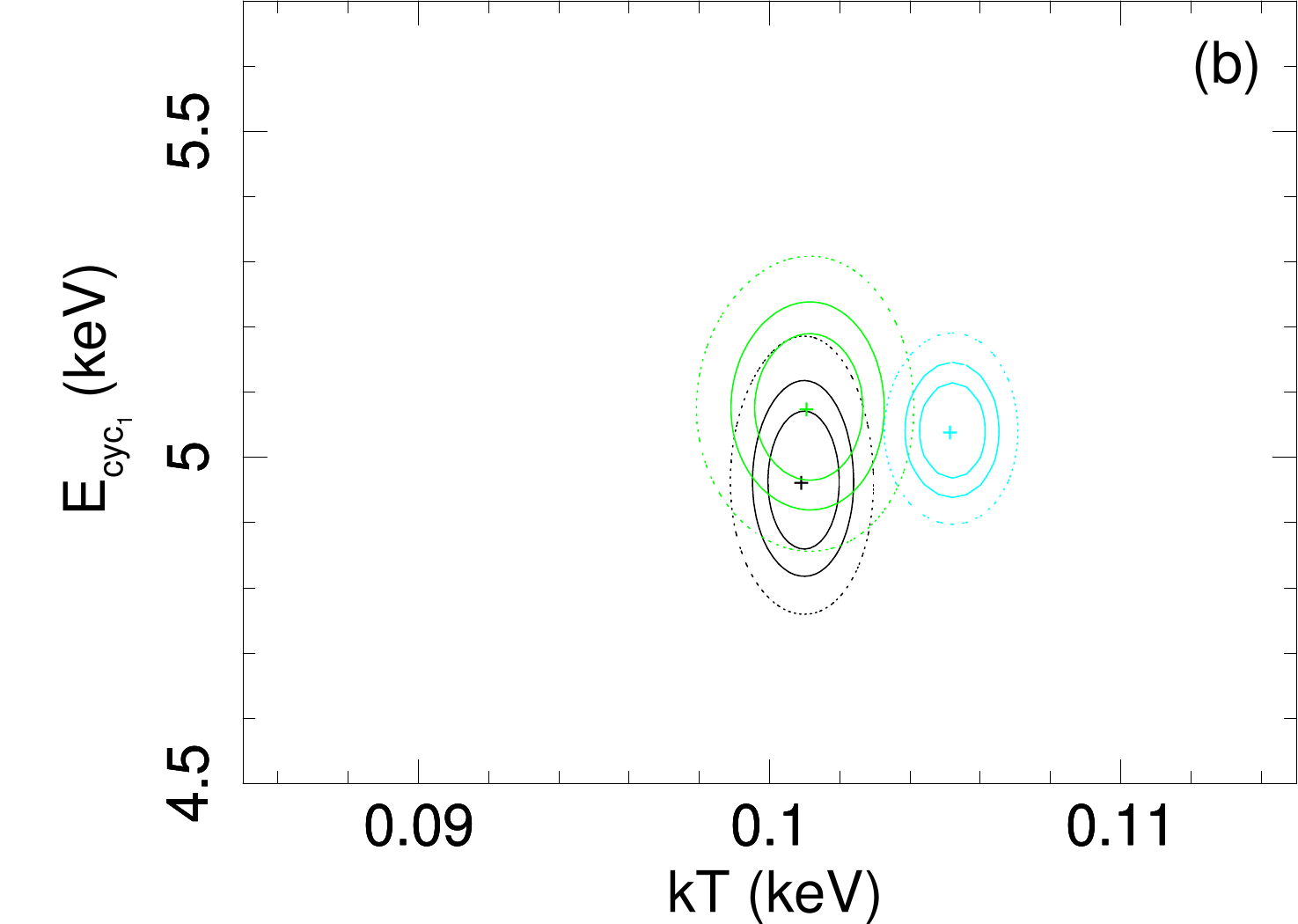}
\end{minipage}
\hspace{0.07\linewidth}
\begin{minipage}{0.3\textwidth}
\includegraphics[height=1.2\columnwidth,angle=-90]{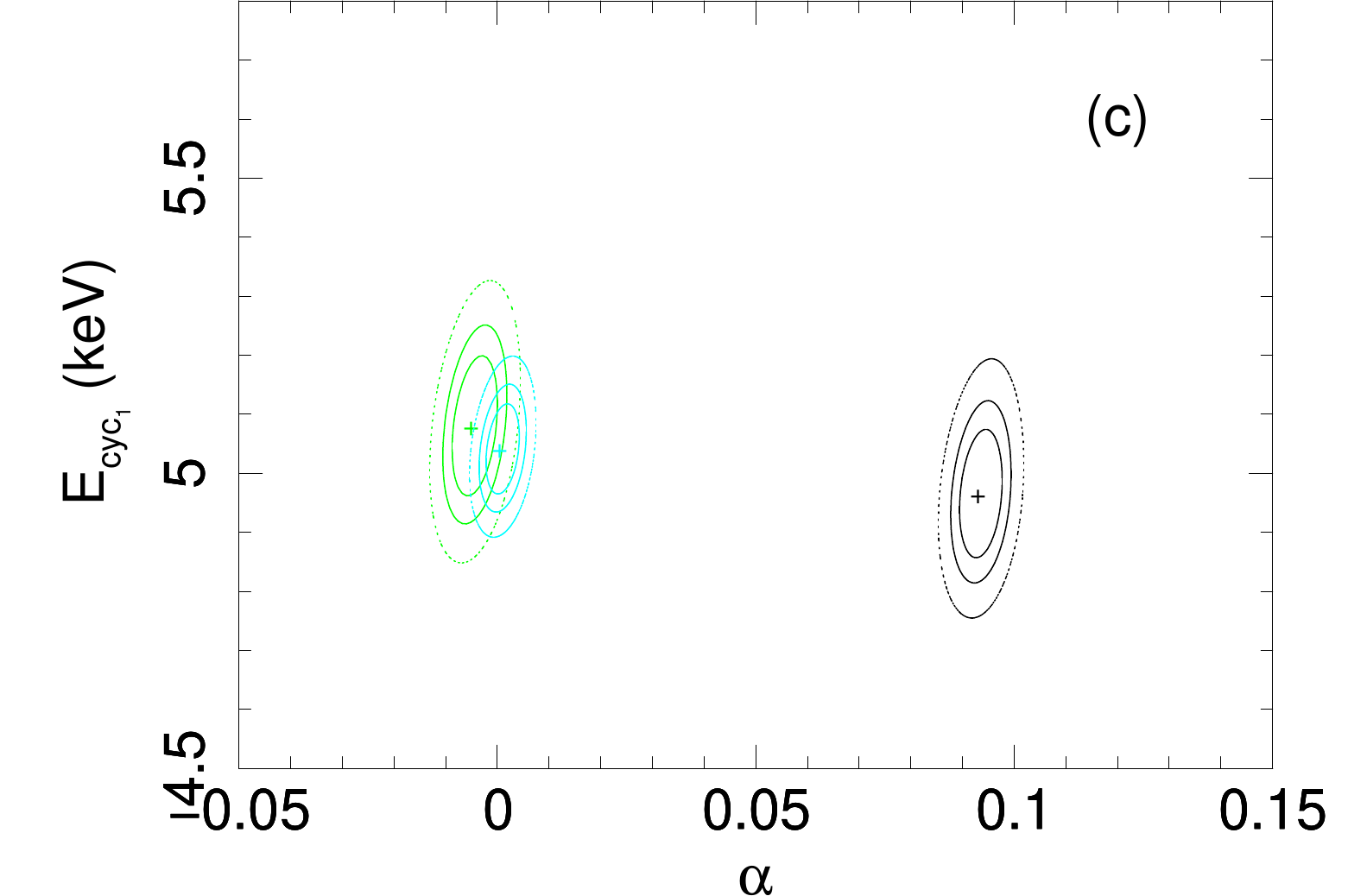}
\end{minipage}
\hspace{0.08\linewidth}
\begin{minipage}{0.3\textwidth}
\includegraphics[height=1.2\columnwidth,angle=-90]{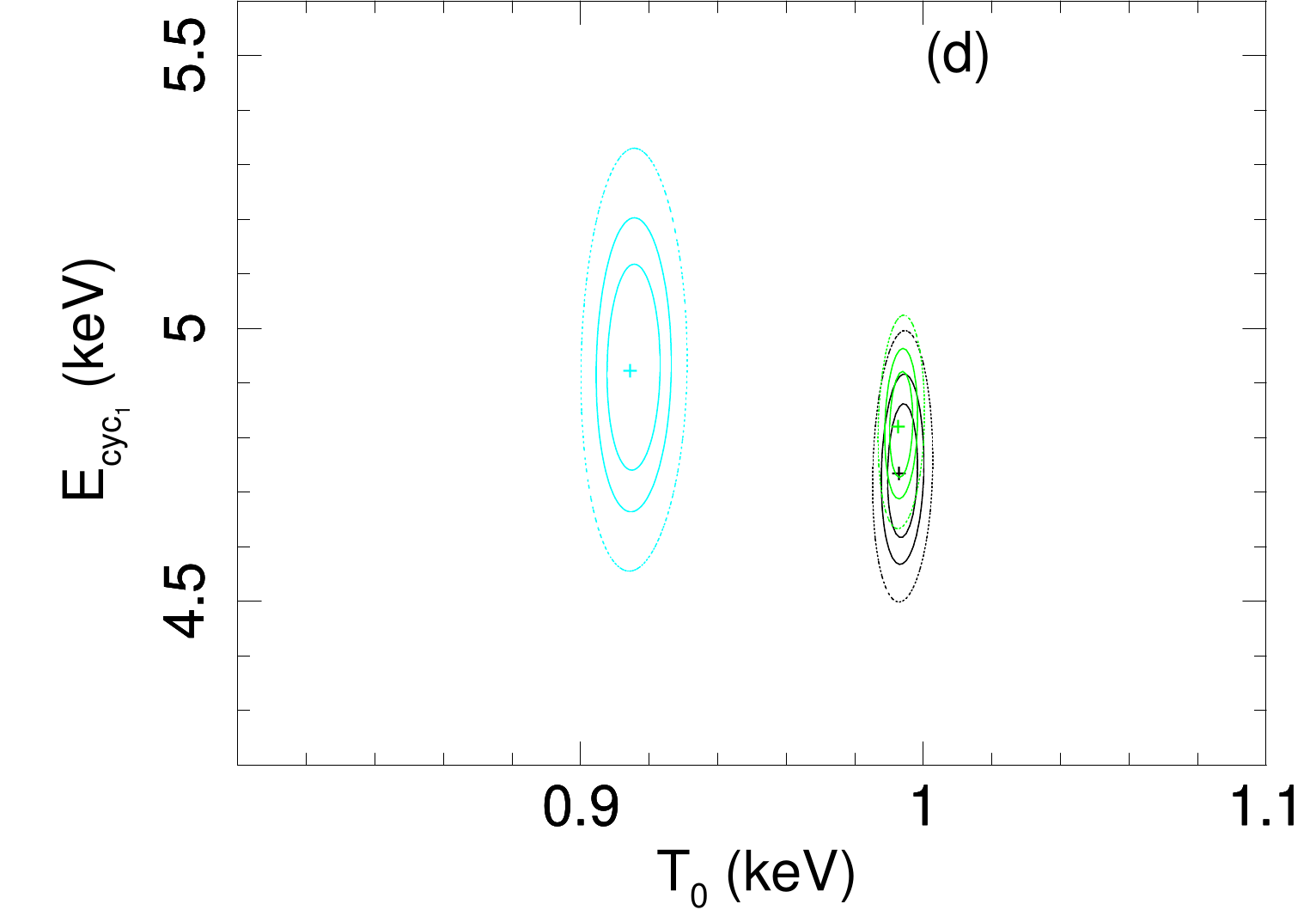}
\end{minipage}
\hspace{0.0465\linewidth}
\begin{minipage}{0.3\textwidth}
\includegraphics[height=1.19\columnwidth,angle=-90]{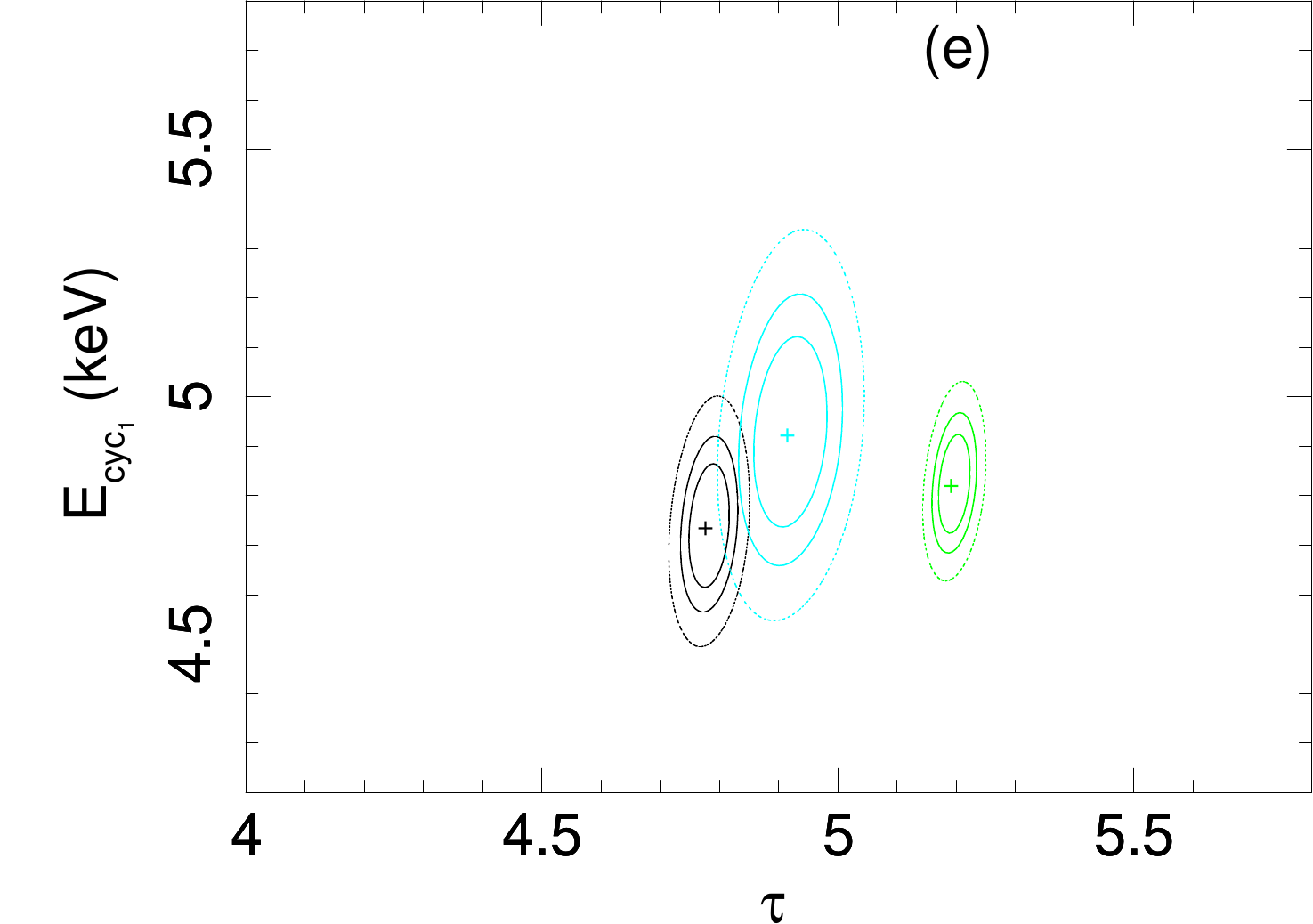}
\end{minipage}
\caption{$\chi^{2}$-contour plots between $E_{cyc_{1}}$ and the other parameters for different observations. The labels (a), (b), (c), (d) and (e) represent the contour plots between $E_{cyc_{1}}$ and $n_{H}$, $kT$, $\alpha$, $T_{0}$ and $\tau$ respectively. Black, green and cyan contours corresponds to observation IDs 4202070101, 4202070102 and 4202070103 respectively. The three contours in each observation corresponds to 68\% (innermost), 90\% (middle) and 95\% (outermost) uncertainty of two parameters of interest. The cross inside the contours represents the best-fit solution. $n_{H}$ value plotted here is obtained when \textsc{cutoffpl} is used as a continuum model. $kT$ is the blackbody temperature of the model \textsc{bbody}, $\alpha$ is the photon-index. $T_{0}$ and $\tau$ are the parameters of the model \textsc{comptt}.}  
\label{Fig-7}
\end{figure*}

In order to rule out the possible artificial dependence of $E_{cyc_{1}}$ on the other model parameters, we plotted $\chi^{2}$-contour between $E_{cyc_{1}}$ and other parameters (Figure \ref{Fig-7}). Most of the contours are found to be almost perpendicular to the one of the axis, and the confidence intervals are found to be well separated. Therefore, we can conclude that there exists either a weak or no dependence of the $E_{cyc_{1}}$ on any other model parameter. Thus, the dependence of $E_{cyc_{1}}$ on luminosity is real and is not due to the dependence of $E_{cyc_{1}}$ on the other model parameters.

\subsection{Variation of Photon-index with flux}
\begin{figure}
\includegraphics[scale=0.35]{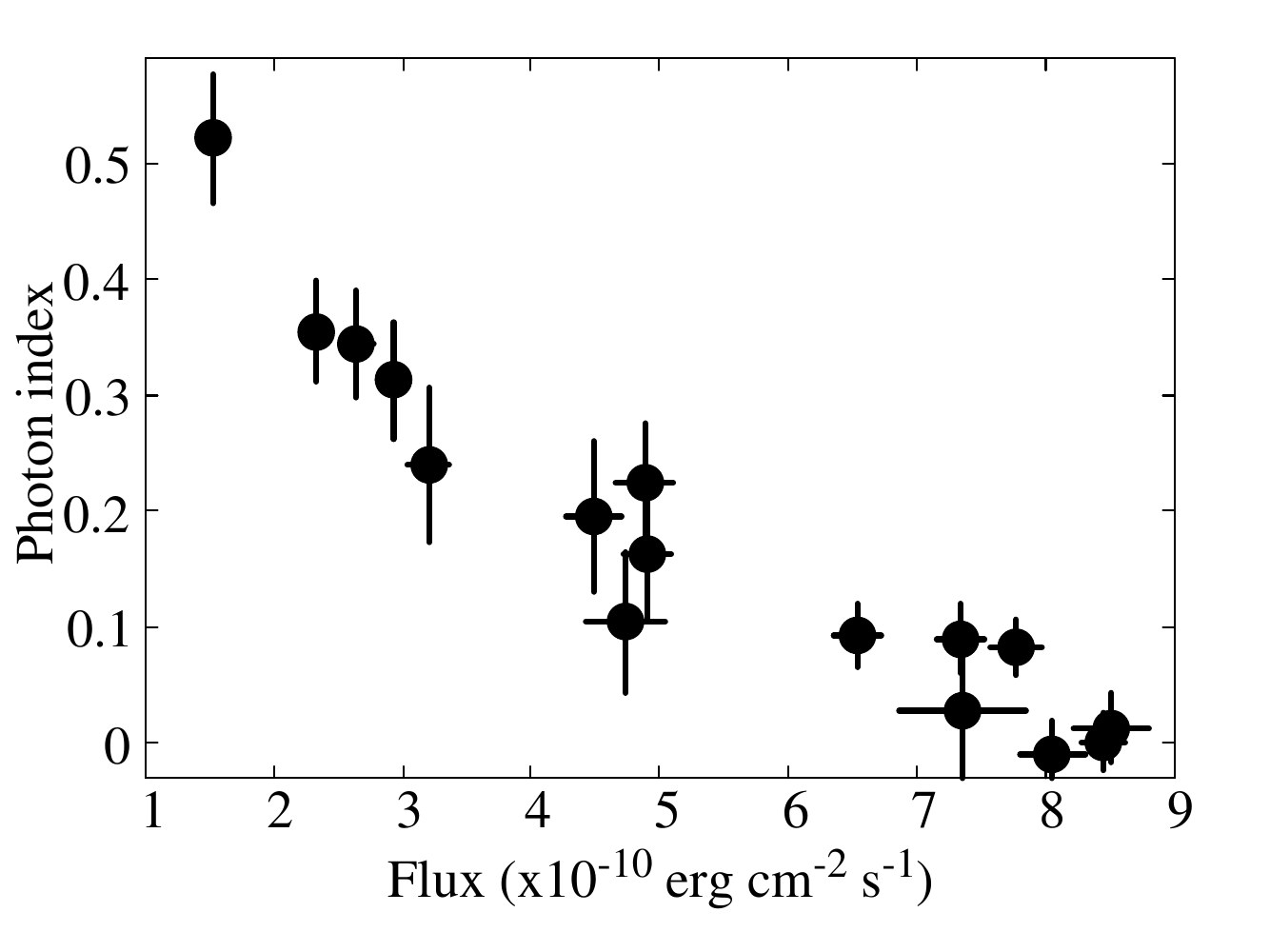}
\caption{Variation of the photon-index with flux.}
\label{fig-8}
\end{figure}

The photon-index gives a measure of softness or hardness of the spectrum. $\alpha$ is high when the flux is low and vice-versa. So there is a decrease in $\alpha$ as flux increases. In other words, as the flux increases the spectrum becomes more harder. This has been shown in Figure \ref{fig-8}.

\section{Discussion}
We have studied the variation of centroid energy of $\sim$ 5 keV ($E_{cyc_{1}}$) cyclotron line of Swift J1626.6-5156 with luminosity $(L_{x})$ using \emph{NICER}. A positive correlation is observed between $E_{cyc_{1}}$-$L_{x}$. In general, the centroid energy of the cyclotron line ($(E_{cyc}$ ) of X-ray pulsars is dependent on the luminosity. The correlation between $E_{cyc}$-$L_{x}$ can be positive or negative depending on the accretion state of the pulsars. Although some X-ray pulsars exist where positive and negative correlations are observed (Cusumano et al. 2016). The accretion state of an accreting X-ray pulsar is separated by a certain value of luminosity known as critical luminosity. The critical luminosity is the minimum value of the luminosity sufficient to stop the accretion of matter above the surface of the neutron star by the radiation pressure (Basko \& Sunyaev 1976; Becker et al. 2012; Mushtukov et al. 2015a). The critical luminosity of an X-ray pulsar depends on its magnetic field (Becker et al. 2012) and is given as,

\begin{equation}
L_{crit}=1.49\times10^{37}\left(\dfrac{B}{10^{12} G}\right)^{16/15} erg\;s^{-1}
\end{equation} 
\label{eq-1}
where $B$ is the magnetic field of the neutron star.

 For the luminosity above the critical value (super-critical state), an anti-correlation between cyclotron line and luminosity is observed, like \textit{in} 4U 0115+63 (Mihara et al. 2004), V0332+53 (Tsygankov et al. 2006) and 1A 0535+262 (Mandal\& Pal 2022). In the sub-critical state (below critical luminosity), a positive correlation between $E_{cyc}$ and $L_{x}$ like in Her X-1 (Staubert et al. 2007), Vela X-1 (Fürst et al. 2014) and 2S 1553-542 (Malacaria et al. 2022). The magnetic field corresponding to the fundamental cyclotron line at $\sim$ 5 keV is $\sim$ 5.6$\times$10$^{11}$ G, assuming a redshift $z=0.3$ on the surface of the neutron star (Maitra et al. 2018). The value of $L_{crit}$ obtained from Equation \ref{eq-1} is $\sim$8.03$\times$10$^{36}$ erg s$^{-1}$. The estimated distance to the source is about 5.8-12 kpc (Bailer-Jones et al. 2021). If we assume \textit{the} distance to the source to be 10 kpc, then most of the observations lie below the critical luminosity. However, due to large uncertainty in the measured distance, the luminosity cannot be calculated precisely, and all the observations may lie in the sub-critical state. The morphology of the pulse profiles suggests that the source is in the sub-critical state. In the sub-critical regime of the accretion, the radiation pressure is not sufficient to stop the matter from flowing onto the surface of the neutron star, however, it can decrease the velocity of infalling matter below the free fall velocity. Cyclotron lines are formed close to the neutron star's surface in the accreting matter through resonant scattering and are redshifted with respect to the actual (local) energy. The amount of redshift depends on the velocity of the matter accreting to the neutron star \textit{i.e.} on the luminosity of the X-ray pulsars (Mushtukov et al. 2015b). In the sub-critical regime of accretion, an increase in luminosity will decrease the velocity of the accreting matter as the radiation pressure will be strong, due to which there will be a small redshift in the centroid energy of the actual cyclotron line. Thus, we will then observe a positive correlation of cyclotron line energy with the luminosity (Mushtukov et al. 2015b). Mushtukov et al. (2015c) provided a more accurate estimation of the critical luminosity by taking resonances of the resonant scattering of the Compton scattering cross-section, polarization and configuration of accretion flow into account. According to their result, the critical luminosity is not a monotonic function of the magnetic field as Becker et al. (2012) obtained but rather shows a complex dependence on the magnetic field. The model successfully explained the observed variation of the cyclotron line with the luminosity of the sources like A 0535+26, V0332+53, 4U 0115+63, and GX 304-1. Recently, Raman et al. (2021) have provided the update $E_{cyc}-L_{crit}$ curve based on the result of Mushtukov et al. (2015c). The $E_{cyc}-L_{crit}$ curve is shown in Figure 7 of Raman et al. (2021), and one can observe that for a $\sim$ 5 keV cyclotron line the critical luminosity is $\sim$ 10$^{37}$ erg s$^{-1}$ thus supporting our argument that the source is in the sub-critical accretion state.

In a sub-critical state, the X-ray emission occurs from a surface or close to the surface of the neutron star (Basko \& Sunyaev 1976). The X-ray emission takes place in a pencil beam pattern and is highly pulsed. The resulting pulse profile is single-peaked. However, the radiation pressure dominates for the luminosity above critical luminosity, and radiation-dominated shock is formed above the neutron star's surface. The X-ray emission takes place from the accretion column's side wall, resulting in a fan beam emission pattern. The pulse profile, in this case, is complex with multiple peaks. The single peaked pulse profiles of Swift J1626.6-5156 during \emph{NICER} observations hint that it is in a sub-critical state. The change in the shape of the pulse profiles at 59315.2 MJD and 59324.6 MJD (Figure \ref{fig-3}) can be due to a decrease in the sound-to-noise ratio in the low-intensity state (Figure \ref{Figure1}).

Figure \ref{fig-8} shows that the photon-index $\alpha$ is high when flux (or luminosity) is low. Thus, the spectrum is softer (higher value of $\alpha$) in the low flux states compared to high flux states. Thus, an anti-correlation exists between $\alpha$ and $L_{x}$. Such anti-correlation between $\alpha$ and $L_{x}$ was observed in Swift J1626.6-5156 by Icdem et al. (2011) and Reig \& Nespoli (2013). Reig \& Nespoli (2013) studied the timing and spectral properties of Be X-ray pulsars like 4U 0115+63, EXO 2030+375, V 0332+53, KS 1947+300, 1A 0535+262, and others during an outburst. They observed that the $\alpha$ is anti-correlated with $L_{x}$ in the sub-critical regime of accretion. Such anti-correlation has also been observed in 2S 1417-624 (Gupta et al. 2018)  and 2S 1553-542 (Rai et al. 2023) during the sub-critical accretion state. Becker et al. (2012) showed that the height of the emission region in the accretion column depends on the luminosity and the accretion state. For the sub-critical accretion state, the height of the emission region decreases with the increase in luminosity (refer Eq. (51) of Becker et al. 2012), whereas in the super-critical accretion state, the height of the emission region increases with an increase in the luminosity (refer Eq. (40) of Becker et al. 2012). In the super-critical state, the diffusion and advection are balanced, due to which the bulk comptonizing electron will not gain enough energy to move in higher energy bands, and hence softening of the photon is expected with an increase in luminosity. An increase in luminosity in the sub-critical state decreases the emission region's height, consequently increasing the plasma's optical depth in the sinking zone. This increase in optical depth results in harder photons (Reig \& Nespoli 2013). So, the observed anti-correlation of photon index with flux may result from the hardening of the spectrum with an increase in flux in the sub-critical accretion state.

 \section*{Data Availability}

 The \emph{NICER} observational data used in this research are downloaded from NASA HEASARC data archive and is available publicly.

\section*{Acknowledgements}

 This research has made use of software and publicly availaible data provided by the High Energy Astrophysics Science Archive Research Center (HEASARC), which is a service of the Astrophysics Science Division at NASA/GSFC. The authors would like to thank ICARD, NBU for the providing extended research facilities. The authors are thankful to the reviewer for his/her valuable suggestions to improve the quality of the paper.











\end{document}